\documentclass[journal,9pt]{IEEEtran}
\usepackage{amsmath,amsfonts}
\usepackage{algorithmic}
\usepackage{algorithm}
\usepackage{array}
\usepackage[caption=false,font=normalsize,labelfont=sf,textfont=sf]{subfig}
\usepackage{textcomp}
\usepackage{stfloats}
\usepackage{url}
\usepackage{verbatim}
\usepackage{graphicx}
\usepackage{cite}
\usepackage{bm}
\usepackage{multirow}
\usepackage{tabularray}
\usepackage{xcolor}

\usepackage{listings}
\usepackage{hyperref}
\hypersetup{colorlinks=false,pdfborder=0 0 0,citebordercolor={0 1 0}}

\newcommand{\textfw}[1]{\scalebox{.8}[1.0]{\texttt{#1}}}

\begin{document}

\title{ONNXim: A Fast, Cycle-level Multi-core NPU Simulator}

\author{Hyungkyu~Ham\IEEEauthorrefmark{1},
Wonhyuk~Yang\IEEEauthorrefmark{1}, 
Yunseon~Shin, 
Okkyun~Woo, 
Guseul~Heo, 
Sangyeop~Lee, 
Jongse~Park, 
Gwangsun~Kim
\thanks{\IEEEauthorrefmark{1}These authors contributed equally to this work.}
\IEEEcompsocitemizethanks{\IEEEcompsocthanksitem H. Ham, W. Yang, Y. Shin, O. Woo, and G. Kim are with the
Department of Computer Science and Engineering, POSTECH, Pohang, Republic of Korea 37673.
Email: \{hhk971, wonhyuk, ysshin, okkyun.w, g.kim\}@postech.ac.kr.}
\IEEEcompsocitemizethanks{\IEEEcompsocthanksitem G. Heo, S. Lee, and J. Park are with the
School of Computing, KAIST, Daejeon, Republic of Korea 34141.
Email: \{gsheo, sangyeop-lee, jspark\}@casys.kaist.ac.kr.}
}


\maketitle
\begin{abstract}
As DNNs are widely adopted in various application domains while
demanding increasingly higher compute and memory requirements, 
designing efficient and performant NPUs (Neural Processing Units) is 
becoming more important. However, existing architectural NPU simulators 
lack support for high-speed simulation, multi-core modeling, 
multi-tenant scenarios, detailed DRAM/NoC modeling, and/or different deep learning frameworks.
To address these limitations, this work proposes \emph{ONNXim}, 
a fast cycle-level simulator for multi-core NPUs in DNN serving systems. 
It takes DNN models represented in the ONNX graph format generated from various
deep learning frameworks for ease of simulation. 
In addition, based on the observation that typical NPU cores  process tensor tiles
from on-chip scratchpad memory with \emph{deterministic} compute latency, 
we forgo a detailed modeling for the computation while still preserving
simulation accuracy. 
ONNXim also preserves dependencies between compute and tile DMAs.
Meanwhile, the DRAM and NoC are modeled in cycle-level
to properly model contention among multiple cores that can execute
different DNN models for multi-tenancy.
Consequently, ONNXim is significantly faster than existing simulators 
(e.g., by up to 384$\times$ over Accel-sim)
and enables various case studies, such as multi-tenant NPUs, that
were previously impractical due to slow speed and/or lack of functionalities.
ONNXim is publicly available at \url{https://github.com/PSAL-POSTECH/ONNXim}.

\end{abstract}

\IEEEpeerreviewmaketitle

\vspace{-.15in}
\section{Introduction}

\IEEEPARstart{A}s the accuracy of Deep Neural Networks (DNNs) rapidly improves, 
they are becoming more widely adopted across various domains
while demanding more compute and memory resources.
Thus, it has become increasingly important to design efficient and high-performance
Neural Processing Units (NPUs) for DNNs. 

Since architectural design exploration relies heavily on simulators, 
various cycle-level NPU simulators have been proposed.
However, some of them, such as STONNE~\cite{stonne}, SCALE-Sim~\cite{ScaleSim}, 
and Timeloop~\cite{timeloop}, lack support for critical aspects of NPUs such as modeling of
multi-core NPUs and/or detailed modeling for important shared resources (e.g., DRAM).
Detailed modeling for shared resources is especially important for server-class NPUs
that require multi-tenancy for high resource utilization~\cite{tpuv4i}.
Other simulators such as Accel-Sim~\cite{Accelsim}\footnote{Although Accel-Sim is not an NPU simulator,
it models the Tensor Core in NVIDIA GPUs and GPUs are currently widely used to accelerate
DNNs. Thus, we include it in our discussion and comparisons.}, SMAUG~\cite{xi2020smaug},
mNPUSim~\cite{hwang2023mnpusim} do support multi-core NPUs and detailed DRAM modeling, but
significantly sacrifice the simulation speed. 
Considering that both DNNs and NPUs are becoming larger, increasing the time complexity of 
the simulation, it is important to provide fast simulation speeds.
While FPGA-based simulation (e.g., Gemmini~\cite{genc2021gemmini} with FireSim~\cite{firesim})
achieves high speed, it cannot be used to model server-class NPUs that do not fit into 
an FPGA. 
In addition, many simulators require model descriptions in special formats
rather than standard formats.
Thus, existing simulators do not meet all of the requirements needed to design
future NPUs as shown in Table~\ref{tab:simulators}.

In this work, we propose ONNXim, a fast, cycle-level NPU simulator that supports
multi-core NPUs, multi-tenancy, and detailed modeling of the shared DRAM and 
NoC resources, to overcome the limitations of existing simulators. 
We leverage ONNX (Open Neural Network Exchange)~\cite{onnx}, an open standard format
for describing deep learning models implemented in different frameworks (e.g.,
PyTorch and TensorFlow), because it is currently one of the most widely used formats 
for DNN model conversion.
For example, ONNX is the recommended input format for TensorRT, which optimizes
inference for NVIDIA GPUs~\cite{TensorRT}.
By using ONNX graphs as the input format, 
our simulator can easily run different DNNs implemented in various frameworks.
ONNXim is recognized as an \emph{execution provider} by the ONNX runtime, 
similar to devices such as CPUs and GPUs, to exploit its graph optimization flow.
It currently supports commonly known operation fusions and can be
easily extended to study the impact of various optimization techniques.

ONNXim achieves high simulation speed based on the observation that typical NPU cores
with systolic arrays process tensor tiles from on-chip scratchpad memory with deterministic
compute latency~\cite{ScaleSim}. 
Since the compute latency can be determined based on the sizes of
the tensor tiles and systolic array, we avoid modeling all individual 
operations in DNNs in a fine-grained manner, unlike conventional 
CPU/GPU simulators, without losing simulation accuracy.
At the same time, shared resources (e.g., DRAM) are modeled at cycle-level since the 
contention from multiple cores introduces non-determinism.  
While a prior work~\cite{hwang2023mnpusim} adopts a similar approach, ONNXim
improves upon it in three significant ways. First, ONNXim enables DNN inferences with dynamic 
input shape, which is important for properly modeling the key-value cache in large language model (LLM) inference
as the sequence length increases during the generation phase. 
Second, in addition to the General Matrix Multiplication (GEMM) and convolution supported in prior work, 
we provide a more diverse set of 
operations, including layer normalization and skip connection, which can collectively take up a significant
portion of runtime~\cite{ghodrati2024tandem}.
Third, in ONNXim, the generation and execution of the dynamic instruction sequence for cores is optimized for fast simulation speed.

We validate the accuracy of our NPU core model against the Gemmini NPU RTL model~\cite{genc2021gemmini}.
For cycle-level models of shared DRAM and NoC, we adopt Ramulator~\cite{Ramulator} and 
Booksim~\cite{booksim}.
We focus on DNN inference and leaves training support as future work.

\begin{table}[]
\caption{Comparison of the features of different simulators.}
\label{tab:simulators}
\scriptsize
\centering
\setlength{\tabcolsep}{2.2pt} %
\begin{tabular}{|l||c|c|c|c|c|c|} 
\hline
&  \bf High &  \bf Multi- &  \bf Multi- &   \bf Detailed &  \bf Optimization &   \bf Model input \\
&  \bf speed &  \bf core &  \bf tenant &   \bf DRAM/NoC &  \bf flow &   \bf format \\ \hline \hline
Accel-Sim~\cite{Accelsim} &  X &  O &  X &  O & O  &  Instr. trace  \\ \hline
mNPUsim~\cite{hwang2023mnpusim} &  X &  O &  O &  O & X  &  Custom  \\ \hline
SCALE-Sim~\cite{ScaleSim} &  O &  X &  X &  X & X  &  Custom \\ \hline
SMAUG~\cite{xi2020smaug} &  X &  O &  X &  O & X &  Custom  \\ \hline
STONNE~\cite{stonne} &  X &  X &  X &  O &  X &  PyTorch   \\ \hline
Timeloop~\cite{timeloop} & O  & X & X  & X  & O &   Custom  \\ \hline
{\bf ONNXim} &  O &  O &  O &  O & O &  ONNX graph \\ \hline
\end{tabular}
\vspace{-.05in}
\end{table}

In summary, the contributions of this work include the following: 
\begin{itemize}
\item ONNXim uses the ONNX graph format for DNN model description to 
run models written in different frameworks 
without any manual conversion. It is integrated with the ONNX runtime 
for the optimization flow over the dynamic computation graph.
\item ONNXim achieves significantly higher simulation speed than 
existing simulators while enabling cycle-level simulation of multi-core NPUs
with accurate shared resource modeling. 
Our NPU core model achieves an average absolute error of 0.23\% over the Gemmini RTL model~\cite{genc2021gemmini}.
\item We demonstrate the usefulness of ONNXim with case studies
on multi-tenant workloads, including LLMs, and the impact of DNN model architectures that
were previously impractical due to slow speed and/or lack of functionalities.
\end{itemize}

\begin{figure*}
    \centering
    \includegraphics[width=.75\linewidth]{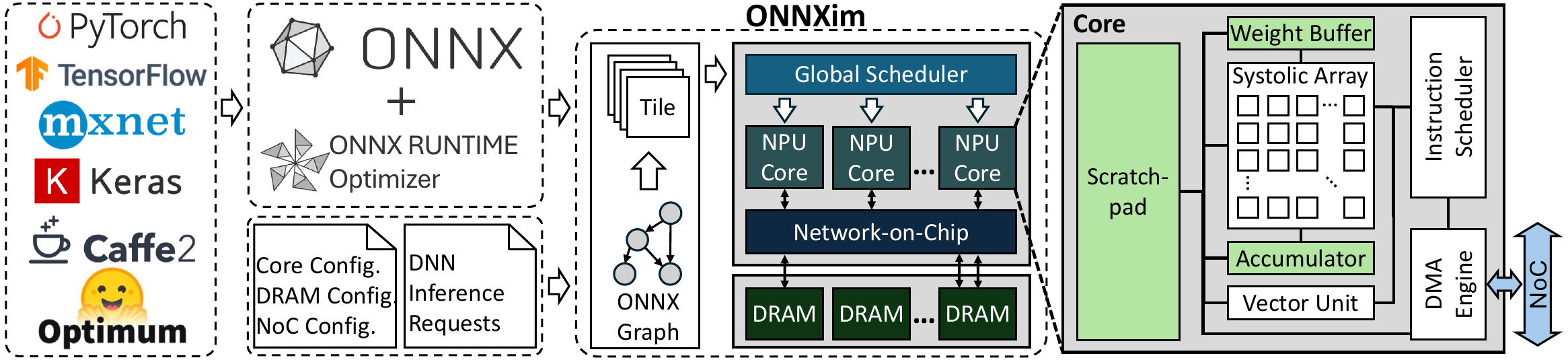}
    \caption{Overview of ONNXim simulator.}
    \label{fig:onnxim_infra}
\end{figure*}

\section{ONNXim}
ONNXim is a cycle-level simulator for inference-oriented multi-core NPUs with
systolic arrays (Fig.~\ref{fig:onnxim_infra}).
It is designed to achieve high simulation speed to enable running large
DNN inference tasks (e.g., LLMs or large batch size)
while accurately modeling contention for shared resource (e.g., DRAM).
For multi-tenant scenarios, it supports multiple scheduling policies
(e.g., time-shared and static) and can be easily extended to support
other policies. 
To enable simulation of various models written in different frameworks 
(e.g., PyTorch and TensorFlow) without manual conversion to a custom format,
ONNXim takes standard ONNX graphs as input model description.
It is implemented as an execution provider (EP) for ONNX runtime along with
other EPs (e.g., CUDA) to exploit its optimization flow.

\subsection{Front End}
\textbf{Model optimization and lowering.}
Using the ONNX runtime's offline optimization tool, the input DNN model is first optimized before simulation. ONNX runtime supports several optimization levels~\cite{ONNXRuntime}.
The \emph{basic} and \emph{extended} optimization levels 
eliminate unnecessary computation and reduce memory traffic 
through constant folding, redundancy elimination, and kernel fusion.
For CNNs, a convolution operation can be fused with a batch normalization and/or a skip connection.
In addition, for transformers, different heads of multi-head attention can be fused and 
a layer normalization can be fused with a skip connection. 

To support multi-tenancy scenarios, it also takes a JSON format input that describes
multiple inference requests with different models, batch sizes, and timestamps. 
The ONNX operations in the DNN's optimized graph are then lowered to tensor tile-level operations
using our tile operation templates.
Dependencies between tile operations are preserved based on the input and output tensors.
The tile sizes are chosen using heuristics from prior work~\cite{genc2021gemmini} that maximizes
the utilization of on-chip scratchpad memory.

\textbf{ISA.}
The tile operation templates use the NPU core's ISA to specify operations using 
following instructions:
1) \textfw{MVIN/MVOUT} for DMA load/store between scratchpad memory and DRAM, 
2) \textfw{GEMM Preload} for loading weights into the systolic array, 
3) \textfw{GEMM} for performing matrix multiplication on the systolic array, 
4) \textfw{IM2COL} for performing image-to-column transformation, 
5) and vector operations (e.g., addition and GELU).
The ISA is an extension of Gemmini's ISA~\cite{genc2021gemmini} with additional instructions for
vector operations and activation functions.

\textbf{Scheduler.}
The global scheduler (Fig.~\ref{fig:onnxim_infra}) keeps track of the dependency between operation nodes 
from the input ONNX graph as well as the status of NPU cores.  
The input graph's operation node is turned into a list of tile-level operations and pushed into 
a \emph{ready tile queue} when its dependency is resolved. 
If there are multiple independent operation nodes in the graph, the tiles from the nodes can be 
pushed into the queue together and executed in parallel across different NPU cores. 
When a core becomes available for a tile operation, the scheduler pops a tile operation from the queue
and issues it to core.

ONNXim provides multiple scheduling policies, including time-sharing and spatial-sharing policies. 
The time-sharing policy schedules a layer from one request at a time before switching to a layer from another request.
Although this policy eliminates resource contention between requests, it may result in resource underutilization as well as unfairness due to differences in layer execution time across models.
The spatial-sharing policy allocates NPU cores among various models, enabling them to execute concurrently. 
However, this approach can lead to performance interference among requests due to contention 
(e.g., DRAM row buffer conflicts). 
ONNXim's scheduler can be easily extended using its interface to add new policies.

\subsection{NPU Microarchitecture Model}

\textbf{Core organization.}
ONNXim models a typical NPU core's organization with a systolic array, weight buffer, 
scratchpad memory, accumulator, and vector units (Fig.~\ref{fig:onnxim_infra}).
The accumulator includes its own SRAM as well as arithmetic units for accumulation operations.
The core receives a tile from the global scheduler and executes the instructions within the tile.
Instruction scheduler issues instructions to the DMA engine or execution pipeline when there is
no structural or data hazard.
Once all instructions are issued, regardless of whether they are completed, core receives a new tile from the scheduler if scratchpad and accumulator resources are available.
In this way, the core can run up to two tiles concurrently in a double-buffered manner.
The scratchpad memory and accumulator memory are each partitioned into two parts
and each partition is alternately assigned to tiles issued from the global scheduler 
to enable double buffering. 
The systolic array and vector unit execute an instruction when all of its operands are ready in the scratchpad memory.
The scratchpad memory is modeled to provide a configured word size in a cycle.

\textbf{Core implementation.}
ONNXim improves the simulation speed for NPU core computation 
by avoiding cycle-by-cycle simulation of the systolic array and the vector unit as in conventional simulators for CPUs and GPUs.
The systolic array can be implemented with different dataflows:
weight-stationary, input-stationary, and output-stationary.
Their relative performance depends on the dimensions of the systolic array and
input tensor sizes~\cite{ScaleSim}.
We assume a weight-stationary systolic array since it is widely used in real NPUs~\cite{tpuv1}.
Then, after the weights are preloaded to the systolic array, 
its compute latency can be calculated as $l+width+hegith-1$, 
where $width$ and $height$ are the dimensions of the systolic array, and
$l$ is the dimension of the input tile shifted to the systolic array over time.
Subsequent operations (e.g., \textfw{MVOUT}) that use the systolic array's output 
in the accumulator SRAM will be issued afterwards.

Vector operations are also assumed to use operands from
the core's on-chip SRAM and result in deterministic execution time.
The execution time is calculated based on the size of the operands,
the width of the vector unit specified in the configuration file,
and the vector operator's type from the DNN model. 
The configuration file also specifies the operation latency 
for each operator type.

\textbf{Shared resources.}
We model the off-chip DRAM shared by all NPU cores using 
Ramulator~\cite{Ramulator}, a fast cycle-level simulator that
can model various DRAM devices (e.g., DDR, HBM, and LPDDR), to 
accurately model memory traffic contention. 
For tensor data movement to/from the cores,
the per-core DMA engine generates
memory requests in the DRAM-access-granularity and sends them 
to memory controllers.
ONNXim uses the IPOLY hash scheme~\cite{rau1991pseudo} to load-balance
different memory channels. 

ONNXim provides two NoC models, a simple NoC model with configurable
latency and bandwidth as well as a cycle-accurate simulator, 
Booksim~\cite{booksim}, which is able to model various 
NoC topologies. 
While such a detailed NoC model may be unnecessary if is assumed
that the on-chip bandwidth is abundant, multi-die NPUs~\cite{simba} with limited 
die-to-die interconnect bandwidth would require an accurate interconnect model. 
By integrating the cycle-level models, ONNXim can accurately 
model contention for shared resources.

\section{Evaluation}
\subsection{Methodology}
We evaluated the simulation speed of ONNXim in comparison to existing simulators for NPU and GPU, including Accel-Sim~\cite{Accelsim},
mNPUsim~\cite{hwang2023mnpusim}, and SMAUG~\cite{xi2020smaug}.
SCALE-Sim~\cite{ScaleSim} and STONNE~\cite{stonne} were not included because they cannot model multi-core NPUs, which 
is essential for modeling server-class NPUs.
ONNXim was evaluated in two versions: one with a simple 
latency-bandwidth NoC model (\textfw{ONNXim-SN}) and the other with
Booksim~\cite{booksim} NoC model (\textfw{ONNXim}). 
In both cases, DRAM was modeled with Ramulator~\cite{Ramulator}.

We used two NPU configurations, a \textfw{Mobile NPU} that is similar to 
Arm's Ethos-U55~\cite{ARM-Ethos-U55} and a \textfw{Server NPU} that is similar to 
Google's TPUv4i~\cite{tpuv4i}, for evaluation (Table~\ref{tab:core_config}).
For comparisons with Accel-Sim, we used NPU-like GPU configurations:
\textfw{Mobile NPU}-like GPU assumed a single GA102 SM at 566 MHz (for similar FLOPS)
and 128~KB L2 cache with a single GDDR6 channel with 8~b bus width;
\textfw{Server NPU}-like GPU assumed 75 GA102 SMs at 1.1 GHz and 138~MB L2 cache with
two HBM2 stacks.

For speed comparisons with existing NPU simulators, we focused on GEMM operations,
since it can dominate the simulation time and some of the simulators did not 
support other operations such as layer normalization and skip connection. 
However, we do provide end-to-end DNN simulation speed comparisons with Accel-sim
for ResNet-50 and GPT-3 Small. 
Since GPT-3 has two distinct phases, (i.e., prompt summarization and token generation)
with different characteristics, we denote them separately as GPT-3(S)
and GPT-3(G), assuming a 512-token prompt and generating 100 tokens.

\subsection{Simulation Speed}

\begin{table}
\centering
\footnotesize
\caption{NPU configurations for evaluation.}
\label{tab:core_config}
\begin{tabular}{|c||c|c|} 
\hline
{\bf Parameter}   & {\bf Mobile NPU}   & {\bf Server NPU} \\ 
\hline
\hline
Core frequency            & 1 GHz           & 1 GHz            \\ 
\hline
Number of core   &   4      &   4     \\
\hline
Systolic array size    & 8$\times$8      & 128$\times$128   \\ 
\hline
Vector unit (16 ALUs/lane) &  8 lanes & 128 lanes \\
\hline
Scratchpad size per core & 64 KB           & 32 MB            \\ 
\hline
Accumulator SRAM size per core  & 16 KB           & 4 MB            \\ 
\hline
Crossbar NoC (64-bit flit)   & 4$\times$2    & 4$\times$16  \\
\hline
DRAM device (tCL, tRCD,      & DDR4 (22, 22, & HBM2 (2 stacks)\\
tRAS, tWR, tRP in ns) & 56, 24, 22) & (7, 7, 17, 8, 7)\\
\hline
DRAM bandwidth       & 12 GB/s         & 614 GB/s        \\
\hline
\end{tabular}
\end{table}

\begin{figure}%
\vspace{0.10in}
\centering
\includegraphics[width=0.915\linewidth]{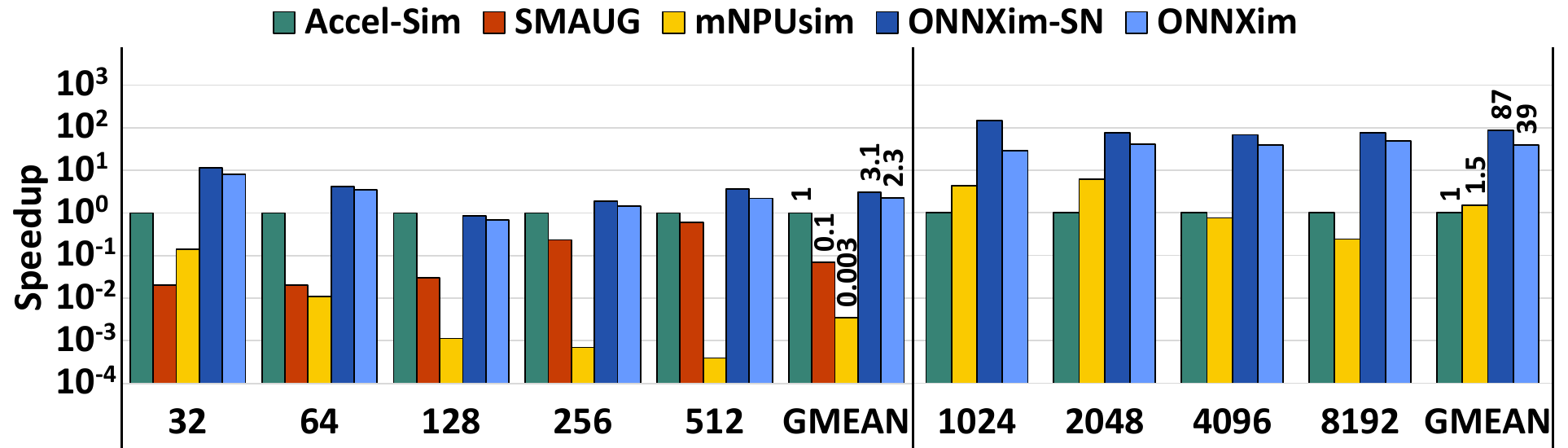}
\hbox{\hspace{0.45in}\footnotesize{(a) \textfw{Mobile NPU}} \hspace{0.7in} \footnotesize{(b) \textfw{Server NPU}}\hspace{0.1in}}
\caption{Comparison of simulation speed over Accel-Sim for GEMM. 
(X-axis: size of each dimension $N$ for $N\times N\times N$ GEMMs.)}
\label{fig:sim_runtime}
\end{figure}

\begin{figure}
    \centering
    \includegraphics[width=0.9\linewidth]{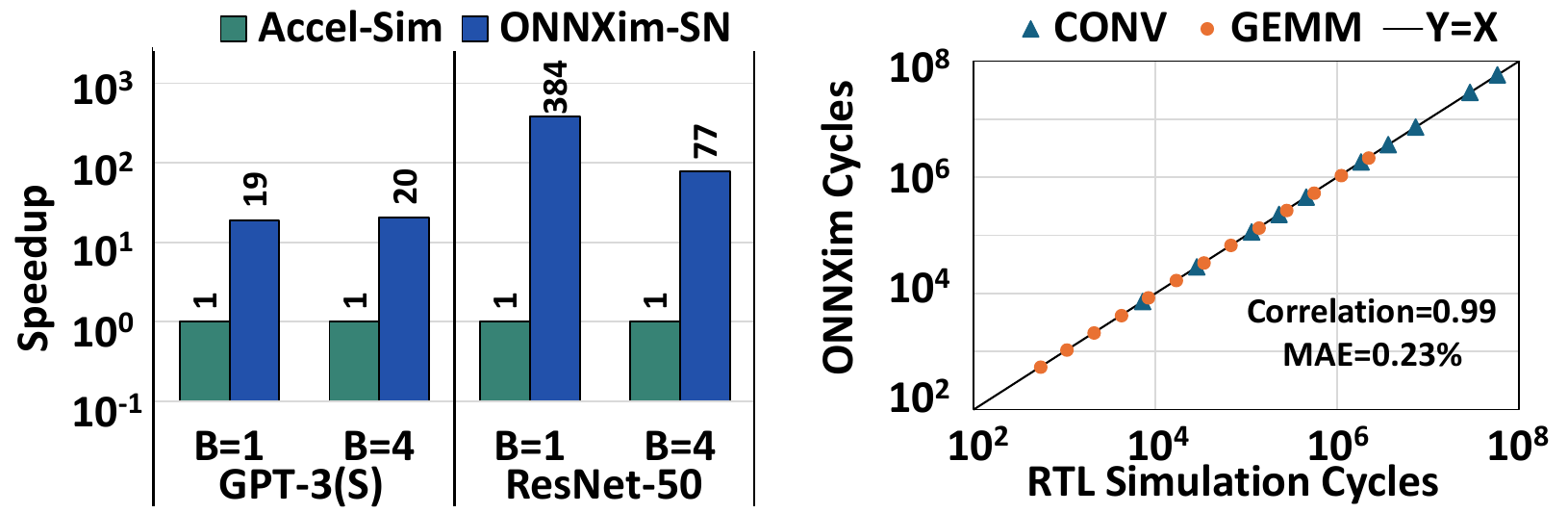} \\
\hbox{\hspace{1.0in}\footnotesize{(a)} \hspace{1.5in} \footnotesize{(b)} \hspace{0.1in}}
    \caption{(a) End-to-end simulation speedup over Accel-sim for different batch sizes (``B'').
    (b) Cycle count comparison between ONNXim and Gemmini RTL model 
    for CONV and GEMM operations for an 8$\times$8 systolic array.}
    \label{fig:e2e_val}
\end{figure}

We used an AMD EPYC 7773X CPU to evaluate the simulation speed. 
All simulators we used, including ONNXim, are single-threaded.
Because SMAUG did not work if the systolic array was larger than 8$\times$8, we used 
the \textfw{Mobile NPU} configuration for the speed comparison. 
For simulation runs that were shorter than two hours, we report average of five measurements.
Compared to Accel-sim, \textfw{ONNXim-SN} achieved significant simulation
speedups of 87$\times$ and 3.1$\times$ for \textfw{Server NPU} and \textfw{Mobile NPU}, 
respectively (Fig.~\ref{fig:sim_runtime}).
While \textfw{ONNXim} provided lower speedup due to the detailed NoC model,
it was still substantially faster than others.
Other simulators were mostly even slower than Accel-sim.
The speedup was pronounced for \textfw{Server NPU}, since as the size of the systolic 
array increases, a larger matrix tile can be processed at once on the simulated NPU core.
In contrast, the number of dynamic instructions in the trace for Accel-sim is proportional to the number of fixed-size tiles from the GEMM.
ONNXim was also significantly faster than Accel-sim in end-to-end DNN simulations.
As shown in Fig.~\ref{fig:e2e_val}a, for the \textfw{Server NPU} configuration running
GPT-3(S) and ResNet-50, \textfw{ONNXim-SN} was 19 to 384 times faster than Accel-sim.

\subsection{Validation with Core RTL Model} \label{validation}

The accuracy of ONNXim's core model was validated against the RTL implementation 
of Gemmini~\cite{genc2021gemmini}. 
While calculating the compute latency of a systolic array is simple, other 
aspects of the core's execution, including data movement between scratchpad memory,
weight buffer, and accumulator as well as the overlap of data movement with systolic array compute latency, should also be accurately modeled and validated. 
We only measured the core's execution time to isolate the randomness from 
memory and NoC latencies. 
The cycle count from ONNXim's core model showed high accuracy 
with the mean absolute error (MAE) of 0.23\% and a correlation of 0.99 
for GEMMs and convolutions of various dimensions (Fig.~\ref{fig:e2e_val}b).

\subsection{Case Study on a Multi-tenant Workload}
\begin{figure}
    \centering
    \includegraphics[width=0.99\linewidth]{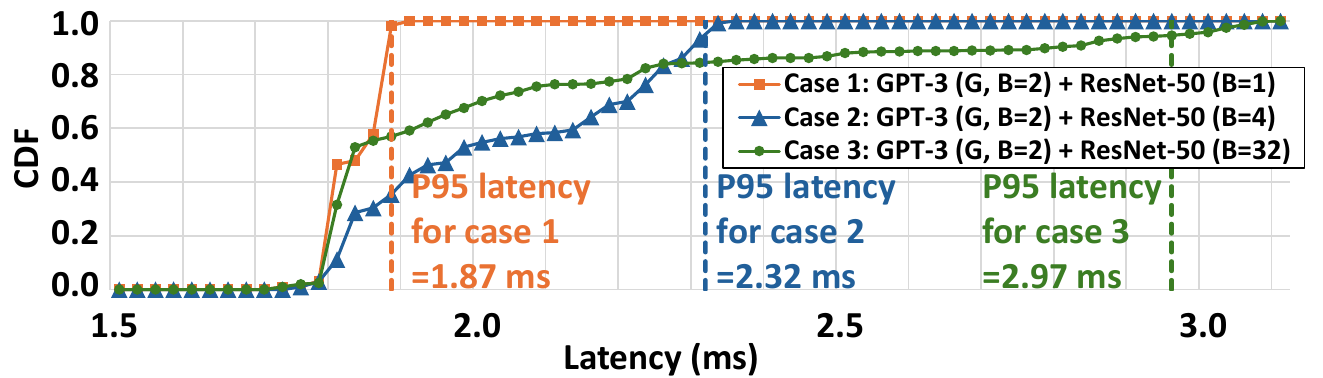}
    \caption{Distribution of TBT for GPT-3's generation phase
    when run with ResNet-50 on the same multi-core NPU with different batch sizes.}
    \label{fig:Latency Distribution}
\end{figure}

For a multi-tenant NPU, the contention for the limited DRAM bandwidth can significantly
impact its performance. As a case study, we show the performance interference from ResNet-50
on the tail latency of GPT-3(G) on a \textfw{Server NPU}. 
The NPU cores were spatially partitioned such that core 0 executed GPT-3(G), 
while cores 1-3 executed ResNet-50.
In Fig.~\ref{fig:Latency Distribution}, we plot the 95th percentile (p95) latency of
TBT (Time-Between-Token) from a total of 500 tokens generated from GPT-3(G) co-executed with
ResNet-50 inferences with different batch sizes. 
The TBT distribution of GPT-3(G) was significantly impacted by the ResNet-50
tasks, increasing the p95 latency by 58\% as the batch size for ResNet-50 varied from
1 to 32.
Using a larger batch for ResNet-50 increases the memory bandwidth demand, resulting
in more contention with the memory traffic from GPT-3(G).
Since inference tasks have latency constraints~\cite{tpuv4i}, request scheduling
has to be done carefully, and ONNXim can be used to study the impact of scheduling 
policies.

\subsection{Case Study on the Impact of Attention Mechanism}
In the generation phase of Transformer-based LLMs,
the attention mechanism can account for a large
portion of inference time. With the multi-head attention (MHA)
from the original Transformer, the query vector of a newly generated token 
is multiplied by the key vector of all previous tokens,
requiring a long GEMV operation.
Since the GEMV operation is memory-bandwidth-bound, the NPU cores are significantly
underutilized. To address this bottleneck, recent LLMs such as Llama-3~\cite{llama3},
replace the MHA with Grouped Query Attention (GQA), where a group of attention 
heads share the same key-value pairs, reducing the size of the GEMV operation.

\begin{figure}
    \centering
    \includegraphics[width=0.9\linewidth]{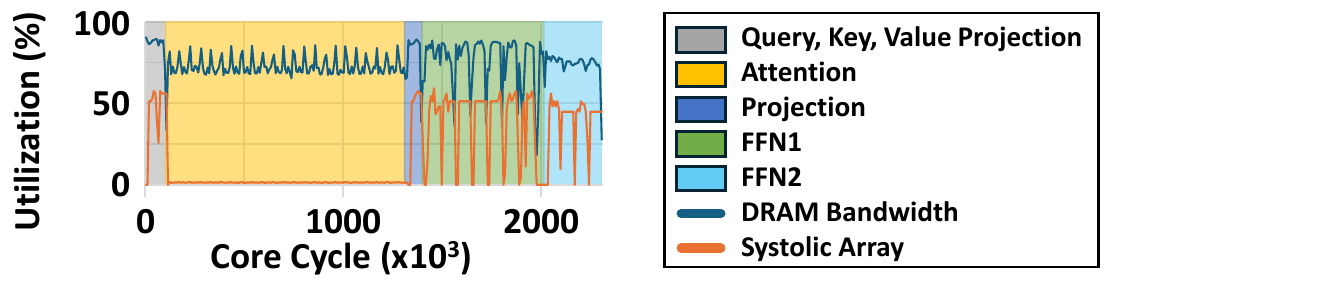}
    \hbox{\footnotesize{(a) Original Llama-3 (8B) with GQA}\hspace{0.8in}}
    \includegraphics[width=0.9\linewidth]{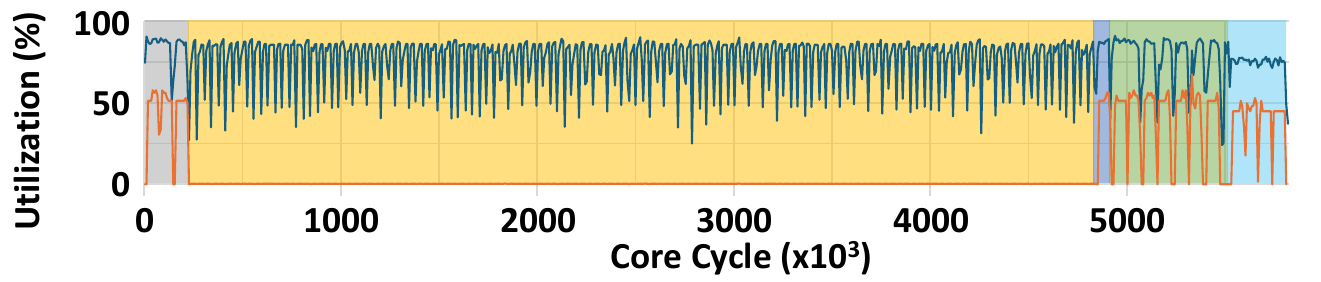}
    \hbox{\hspace{0.1in}\footnotesize{(b) Modified Llama-3 (8B) that replaces GQA with MHA}}
    \caption{Impact of different attention mechanisms on resource utilization.}
    \label{fig:Llama3}
\end{figure}

We studied the impact of the attention mechanism choice on the inference time and
resource utilization in Fig.~\ref{fig:Llama3}. Compared to the original Llama-3 with
GQA, the modified Llama-3 with MHA substantially increases the latency of the attention
mechanism which is memory-bound and underutilizes the NPU core. 
The simulations took 17 and 45 minutes for the two cases on a single CPU core,
which is acceptable for many studies, given the large size of the LLM with 8 billion 
parameters, long context length of 1023 tokens and a large batch size of 128.
This result demonstrates that, 
thanks to the detailed DRAM model and fast simulation speed,
ONNXim can properly model the key characteristics of LLMs
and enables studies of LLM architecture on future NPU designs, 
and vice versa.

\section{Conclusion}

In this work, we introduce ONNXim, a cycle-level multi-core NPU simulator with high simulation speed and 
detailed modeling of shared resources (i.e., DRAM and NoC). 
By leveraging the industry-standard ONNX format for the simulator's DNN model inputs,
ONNXim supports the simulation of existing DNNs implemented in various deep learning frameworks
without manual conversion to a custom format. 
Based on the observation that the compute latency of systolic array can be determined from
its dimensions and the input tile sizes, ONNXim achieves high simulation speedups of up to 384$\times$
compared to prior cycle-level NPU simulators.
Our case studies also show that ONNXim can be used to study various scenarios, 
including multi-tenant DNN inference and HW-aware LLM architecture designs.

\bibliographystyle{IEEEtran}

\begin{thebibliography}{10}
\providecommand{\url}[1]{#1}
\csname url@samestyle\endcsname
\providecommand{\newblock}{\relax}
\providecommand{\bibinfo}[2]{#2}
\providecommand{\BIBentrySTDinterwordspacing}{\spaceskip=0pt\relax}
\providecommand{\BIBentryALTinterwordstretchfactor}{4}
\providecommand{\BIBentryALTinterwordspacing}{\spaceskip=\fontdimen2\font plus
\BIBentryALTinterwordstretchfactor\fontdimen3\font minus \fontdimen4\font\relax}
\providecommand{\BIBforeignlanguage}[2]{{%
\expandafter\ifx\csname l@#1\endcsname\relax
\typeout{** WARNING: IEEEtran.bst: No hyphenation pattern has been}%
\typeout{** loaded for the language `#1'. Using the pattern for}%
\typeout{** the default language instead.}%
\else
\language=\csname l@#1\endcsname
\fi
#2}}
\providecommand{\BIBdecl}{\relax}
\BIBdecl

\bibitem{stonne}
F.~Muñoz-Martínez \emph{et~al.}, ``Stonne: Enabling cycle-level microarchitectural simulation for dnn inference accelerators,'' \emph{IEEE Computer Architecture Letters}, vol.~20, no.~2, pp. 122--125, 2021.

\bibitem{ScaleSim}
A.~Samajdar \emph{et~al.}, ``A systematic methodology for characterizing scalability of dnn accelerators using scale-sim,'' in \emph{IEEE Int'l Symp. on Performance Analysis of Systems and Software}, 2020, pp. 58--68.

\bibitem{timeloop}
A.~Parashar \emph{et~al.}, ``Timeloop: A systematic approach to dnn accelerator evaluation,'' in \emph{IEEE Int'l Symp. on performance analysis of systems and software}.\hskip 1em plus 0.5em minus 0.4em\relax IEEE, 2019, pp. 304--315.

\bibitem{tpuv4i}
N.~P. Jouppi \emph{et~al.}, ``Ten lessons from three generations shaped google’s tpuv4i: Industrial product,'' in \emph{ACM/IEEE 48th Int'l Symp. on Computer Architecture (ISCA)}.\hskip 1em plus 0.5em minus 0.4em\relax IEEE, 2021, pp. 1--14.

\bibitem{Accelsim}
M.~Khairy \emph{et~al.}, ``Accel-sim: An extensible simulation framework for validated gpu modeling,'' in \emph{2020 ACM/IEEE 47th Annual International Symposium on Computer Architecture (ISCA)}, 2020, pp. 473--486.

\bibitem{xi2020smaug}
S.~Xi \emph{et~al.}, ``Smaug: End-to-end full-stack simulation infrastructure for deep learning workloads,'' \emph{ACM Transactions on Architecture and Code Optimization (TACO)}, vol.~17, no.~4, pp. 1--26, 2020.

\bibitem{hwang2023mnpusim}
S.~Hwang \emph{et~al.}, ``mnpusim: Evaluating the effect of sharing resources in multi-core npus,'' in \emph{2023 IEEE International Symposium on Workload Characterization (IISWC)}.\hskip 1em plus 0.5em minus 0.4em\relax IEEE, 2023, pp. 167--179.

\bibitem{genc2021gemmini}
H.~Genc \emph{et~al.}, ``Gemmini: Enabling systematic deep-learning architecture evaluation via full-stack integration,'' in \emph{2021 58th ACM/IEEE Design Automation Conference (DAC)}.\hskip 1em plus 0.5em minus 0.4em\relax IEEE, 2021, pp. 769--774.

\bibitem{firesim}
S.~Karandikar \emph{et~al.}, ``Firesim: Fpga-accelerated cycle-exact scale-out system simulation in the public cloud,'' in \emph{ACM/IEEE 45th Int'l Symp. on Computer Architecture (ISCA)}.\hskip 1em plus 0.5em minus 0.4em\relax IEEE, 2018, pp. 29--42.

\bibitem{onnx}
\BIBentryALTinterwordspacing
Introduction to {ONNX}. [Online]. Available: \url{https://onnx.ai/onnx/intro/}
\BIBentrySTDinterwordspacing

\bibitem{TensorRT}
\BIBentryALTinterwordspacing
NVIDIA. Nvidia tensorrt. [Online]. Available: \url{https://docs.nvidia.com/deeplearning/tensorrt/quick-start-guide/index.html}
\BIBentrySTDinterwordspacing

\bibitem{ghodrati2024tandem}
S.~Ghodrati \emph{et~al.}, ``Tandem processor: Grappling with emerging operators in neural networks,'' in \emph{Proc. of the 29th ACM Int'l Conf. on Architectural Support for Programming Languages and Operating Systems, Volume 2}, 2024, pp. 1165--1182.

\bibitem{Ramulator}
Y.~Kim \emph{et~al.}, ``Ramulator: A fast and extensible dram simulator,'' \emph{IEEE Computer Architecture Letters}, vol.~15, no.~1, pp. 45--49, 2016.

\bibitem{booksim}
N.~Jiang \emph{et~al.}, ``A detailed and flexible cycle-accurate network-on-chip simulator,'' in \emph{IEEE Int'l Symp. on Perform. Anal. of Syst. and Softw.}, 2013, pp. 86--96.

\bibitem{ONNXRuntime}
\BIBentryALTinterwordspacing
ONNXRuntime. Onnx runtime graph optimization. [Online]. Available: \url{https://onnxruntime.ai/docs/performance/model-optimizations/}
\BIBentrySTDinterwordspacing

\bibitem{tpuv1}
N.~P. Jouppi \emph{et~al.}, ``In-datacenter performance analysis of a tensor processing unit,'' in \emph{Proceedings of the 44th annual international symposium on computer architecture}, 2017, pp. 1--12.

\bibitem{rau1991pseudo}
B.~R. Rau, ``Pseudo-randomly interleaved memory,'' in \emph{Proc. of the 18th Annu. Int'l Symp. on Computer Architecture}, 1991, pp. 74--83.

\bibitem{simba}
Y.~S. Shao \emph{et~al.}, ``Simba: Scaling deep-learning inference with multi-chip-module-based architecture,'' in \emph{Proc. of the 52nd Annu. IEEE/ACM Int'l Symp. on Microarchitecture}, 2019, pp. 14--27.

\bibitem{ARM-Ethos-U55}
\BIBentryALTinterwordspacing
ARM. Ethos-u55. [Online]. Available: \url{https://www.arm.com/products/silicon-ip-cpu/ethos/ethos-u55}
\BIBentrySTDinterwordspacing

\bibitem{llama3}
\BIBentryALTinterwordspacing
Meta llama 3. [Online]. Available: \url{https://llama.meta.com/llama3/}
\BIBentrySTDinterwordspacing

\end{thebibliography}

\vfill

\end{document}